\documentclass[aps,prl,onecolumn, preprint,superscriptaddress,showpacs]{revtex4-1}
\usepackage{amsmath}
\usepackage{mathtools}
\usepackage{graphicx}
\usepackage{subfigure}
\usepackage{float}
\usepackage{epsfig}
\begin{document}

\title{\textbf{Optimization of the Combined Proton Acceleration Regime with a Target Composition Scheme}}

\author{W. P. Yao}
\affiliation{Graduate School, China Academy of Engineering Physics, Beijing 100088, China}

\author{B. W. Li}
\email{li\_baiwen@iapcm.ac.cn}
\affiliation{Institute of Applied Physics and Computational Mathematics, Beijing 100088, China}


\author{C. Y. Zheng}
\affiliation{Institute of Applied Physics and Computational Mathematics, Beijing 100088, China}
\affiliation{Key Laboratory of HEDP of Ministry of Education, CAPT, Peking University, Beijing 100871, China}

\author{Z. J. Liu}
\affiliation{Institute of Applied Physics and Computational Mathematics, Beijing 100088, China}
\affiliation{Key Laboratory of HEDP of Ministry of Education, CAPT, Peking University, Beijing 100871, China}

\author{X. Q. Yan}
\affiliation{Key Laboratory of HEDP of Ministry of Education, CAPT, Peking University, Beijing 100871, China}
\affiliation{State Key Laboratory of Nuclear Physics and Technology, Peking University, Beijing 100871, China}
\affiliation{International Centre for Zetta- Exawatt Science and Technology (IZEST)}

\date{\today}

\begin{abstract}

A target composition scheme to optimize the combined proton acceleration regime is presented and verified by two-dimensional particle-in-cell (2D PIC) simulations by using an ultra-intense circularly-polarized (CP) laser pulse irradiating an overdense hydrocarbon (CH) target, instead of a pure hydrogen (H) one. 
The combined acceleration regime is a two-stage proton acceleration scheme combining the radiation pressure dominated acceleration (RPDA) stage and the laser wakefield acceleration (LWFA) stage sequentially together.
With an ultra-intense CP laser pulse irradiating an overdense CH target, followed by an underdense tritium plasma gas, protons with higher energies (from about $20$ GeV up to about $30$ GeV) and lower energy spreads (from about $18\%$ down to about $5\%$ in full-width at half-maximum, or FWHM) are generated, as compared to the use of a pure H target.
It is because protons can be more stably pre-accelerated in the first RPDA stage when using CH targets.
With the increase of the carbon-to-hydrogen density ratio, the energy spread is lower and the maximum proton energy is higher.
It also shows that for the same laser intensity around $10^{22}$ $Wcm^{-2}$, using the CH target will lead to a higher proton energy, as compared to the use of a pure H target.
Additionally, proton energy can be further increased by employing a longitudinally negative gradient of a background plasma density.

\end{abstract}
\pacs{52.38.Kd, 41.75.Jv, 52.35.Mw,52.59.-f}
 \maketitle

\section{Introduction}

Proton acceleration from laser-plasma interaction has been of great interest to researchers over the past few decades \cite{experiments1, experiments2, experiments3} because of its importance for a low-cost tabletop accelerators \cite{accelerator}, fast ignition in inertial confinement fusion \cite{fastignition}, high resolution radiographing \cite{imaging}, cancer radiation therapy \cite{cancer1, cancer2} and laboratory astrophysics \cite{astro}.
Radiation pressure dominated acceleration (RPDA) \cite{pukhov2001, bfshen2001, esirkepov2004, fuchs2006, hegelich2006, yin2007, yan2008, robinson2008, oklimo2008, qiao2009,henig2009,macchi2009} has been proposed as an effective mechanism to obtain hundreds of MeV of monoenergetic and collimated proton beams. Unfortunately, RPDA suffers from multi-dimensional effects, such as the target deformation and the transverse "Rayleigh-Taylor-like" instability. The acceleration process cannot be as stable as predicted by theory or 1D PIC simulations, which results in a shortening of the acceleration length, a reduction of maximum energy and a broadening of the energy spectrum of the proton beam. 

As a result, a lot of attempts have been made to suppress those multi-dimensional effects, among which are, the laser mode scheme \cite{bulanov2008, mchen2008, mlzhou2014}, shaped target scheme \cite{mchen2009, zmzhang2012} and target composition scheme \cite{tpyu2010, qiao2010, tpyu2011, qiao2011, qiao2011pop, sinha2012, liu2013}, etc. Specifically speaking, with the optimized laser pulse (like the transverse super-Gaussian pulse or the Laguerre-Gaussian pulse) and the corresponding shaped target structure, the target deformation problem can be prevented, effectively. 
Despite all these efforts mentioned above, it is still difficult to obtain ultra-energetic protons, for instance, of the order of tens of GeV or even higher. So, it is necessary to look for an alternative long-distance and stable proton acceleration mechanism.  

It is known that electrons can be accelerated stably over a very long distance in a plasma wakefield \cite{tajima1979, wlu2006,leemans2006}.
This laser wakefield acceleration (LWFA) regime was proved to be theoretically capable to accelerate protons as long as they can get trapped by the fast-moving wakefield \cite{pukhov2004}. 
Further, a more specific theory about the trapping condition of protons was established by Shen and his group \cite{bfshen2007, bfshen2009} .
Generally speaking, a proton pre-acceleration stage, which is used to increase the proton velocity, and a high plasma gas density, which is used to decrease the wakefield velocity, are needed to satisfy that trapping condition.   
Above theory has been verified by 2D PIC simulations and optimized by adding a mass-limited target in the plasma gas region to improve the monochromaticity of the ultra-energetic proton beams, by combining LWFA with RPDA.

The combined acceleration regime has been further developed in many studies by using optimized laser modes \cite{zhang2010, yu2010, zheng2012} and modulated underdense plasma gas \cite{bake2012, zheng2013}, in order to get ultra-energetic and collimated proton beams. 
Most recently, to further improve the energy spectrum of the proton beams, the target composition scheme has also been proposed and verified by 1D PIC simulations \cite{yao2014}. 
With the use of a multi-component target, on one hand, the heavy ions can supply excess co-moving electrons for protons, provide them with a negative-gradient "bunching" electrostatic field and preserve their stable acceleration. On the other hand, the heavy ion layer will be separated from the lighter proton layer, because of its smaller charge-to-mass ratio, and will expand in space. The expansion of the heavy ion layer behind the proton layer will protect it from the "Rayleigh-Taylor-like" instability.
In such target composition scheme, the pure H target is replaced by a CH target to ensure that a more stable RPDA stage can be realized. With a more stable first stage, the protons in the CH target can be more efficiently  injected into the wakefield than in the pure H target case. Thus, improvement of the monochromaticity of the proton beams can be achieved after the second LWFA stage. 

In this paper, multi-dimensional effects in above target composition scheme in combined proton acceleration regime are investigated and verified by 2D PIC simulations. 
It is found, in 2D cases, that using a CH target can not only lower the energy spread of the proton beam, but can also increase the maximum proton energy, as compared to the use of a pure H target. 
Additionally, it is worth mentioning that the combined regime using H target requires an ultra-high laser intensity to get protons pre-accelerated strong enough by RPDA in the first stage, so that they can get trapped by the fast-moving wakefield in the second stage and be further accelerated to ultra-high energy.
By replacing the pure H target with a CH one, this high requirement for the laser intensity can be eased down.
Moreover, the proton energy can be further increased by employing a longitudinally negative gradient of the background plasma density. 

\section{simulation model}
A two-dimensional spatial and three-velocity fully relativistic electromagnetic particle-in-cell (2D3V PIC) code KLAP \cite{sheng2002,sheng2008} is used to verify this optimized combined proton acceleration regime. 
The total simulation box is $80\lambda(y) \times 50\lambda(z)$ with $\lambda=1$ $\mu m$, which is the wavelength of the laser pulse, corresponding to a $1000\times2000$ cells moving window. 

\begin{figure}[!htb]
\centering
\includegraphics[width=0.8\textwidth]{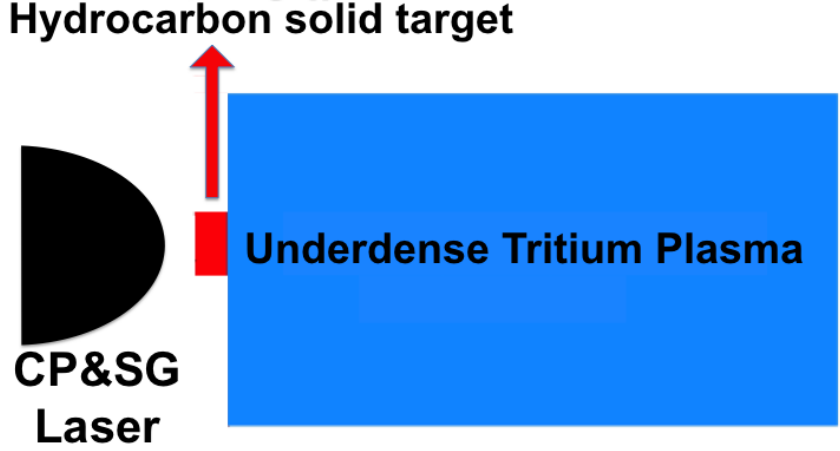}
\caption{(Color online)  The model of the combined proton acceleration regime. The initial density of the hydrocarbon target is $n_e/n_c = 50$, the thickness is $1\mu m$ and the transverse size is $4\mu m$. The density of the background plasma is $n_e/n_c = 0.25$, the length is $600\mu m$ and the transverse size is $80\mu m$. The tritium background plasma has an effective Z/A ratio of the order of 1/3. A circular polarized (CP) laser pulse with a transverse super-Gaussian (SG) profile enters the simulation box from the left boundary.}
\label{fig.1}
\end{figure}

The model is shown in Fig.\ref{fig.1}. A laser pulse is injected initially and propagates along z-axis from the left of the simulation box.  
The normalized amplitude of the circularly polarized (CP) laser amplitude is $a=a_0exp(-r/r_0)^{\kappa}sin^2(\pi t/t_0)$, with $a_0=eA/m_ec^2,r_0=10\lambda, \kappa = 4$ (super-Gaussian, transversely) \cite{zhang2010} and $t_0 = 20\tau$, where $A$ is the vector potential, $c$ is the vacuum light speed, $e$ and $m_e$ are the charge and mass of electrons, and $\tau$ is the laser period, respectively. 

$512$ particles per cell are used in the overdense regions ($10 \lambda < z < 11\lambda, -2\lambda < y < 2\lambda$) and $16$ particles in the underdense plasma gas region ($11\lambda < z < 600\lambda, -40\lambda < y < 40\lambda$). 
The density of the mass-limited overdense target is $n_i = n_e = 50n_c$ (corresponding to $5.5\times10^{22}$ $cm^{-3}$ with $\lambda=1$ $\mu m$), where $n_i = n_p + Z n_C$, $n_p$ is the density of protons, $n_C$ of carbon ions and $n_c$ is the critical density. $Z=6$ as we assume that carbon ions are fully ionized in such an intense laser pulse. Different ratio of $n_C$ to $n_p$ will be examined.
For an underdense plasma gas, the density is $n_i = n_e = 0.25n_c$ (corresponding to $2.8\times10^{20}$ $cm^{-3}$). The density of this plasma gas is higher than in previous works \cite{bfshen2007, bfshen2009, zhang2010, yu2010, zheng2012}, in order to ensure a formation of a slower and more stable wakefield, 
so that protons can be injected into the wakefield more efficiently. 
The tritium plasma gas with an effective Z/A ratio of the order of 1/3 is chosen for the same reason.

\section{simulation results}

\begin{figure}[!htb]
\centering
\includegraphics[width=0.8\textwidth]{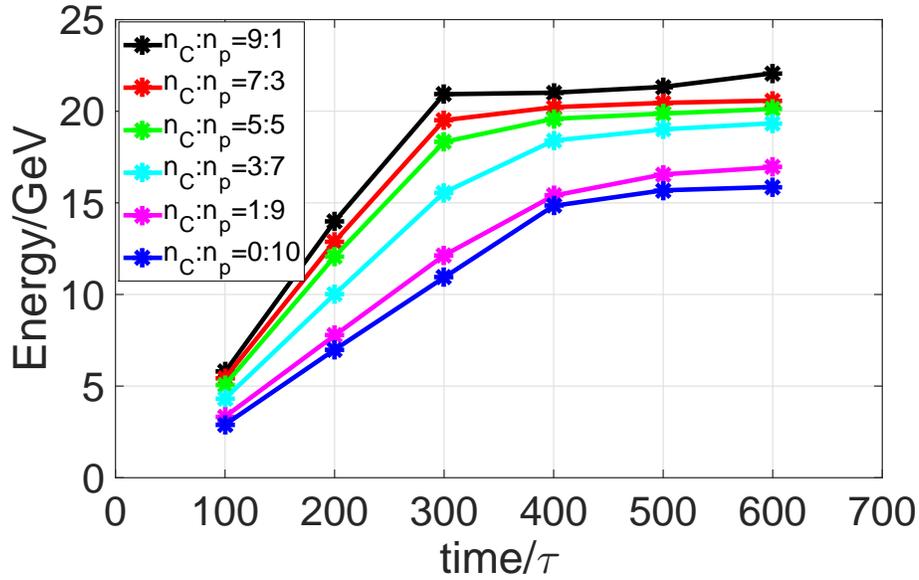}
\caption{(Color online)  Protons within $-10\lambda < y < 10\lambda$ and $30\lambda < z < 50\lambda$ in the moving window are calculated for the average energy. Different density ratios of carbon to proton are shown in different color lines. The blue line ($n_C:n_p=0:10$) means the pure H target, while the black, red, green, cyan and magenta lines are for CH targets with different carbon density ratios; $a_0=200$ with all other parameters are given in the simulation model.}
\label{fig.2}
\end{figure}

Fig.\ref{fig.2} shows the average energy of the protons in $-10\lambda < y < 10\lambda$ and $30\lambda < z < 50\lambda$ in the moving window, evolving with time from $100\tau$ to $600\tau$ in the LWFA stage. 
It is clear that using CH targets can improve the average energy of the protons, effectively. 
Specifically, for the pure H case, proton energy gets saturated after $t=400\tau$, with about $15$ GeV maximum energy. However, for CH cases, the proton energies get saturated earlier than in the H case, at $t=300\tau$ with about $20$ GeV maximum energy, thus increasing the proton average energy by about $33\%$. 
This also indicates that the acceleration efficiency of CH targets is higher than that of H target. 
Additionally, by increasing the percentage of the carbon ions density, the averaged energy and the effectiveness will increase, as well.  

\begin{figure}[!htb]
\centering
\includegraphics[width=0.49\textwidth]{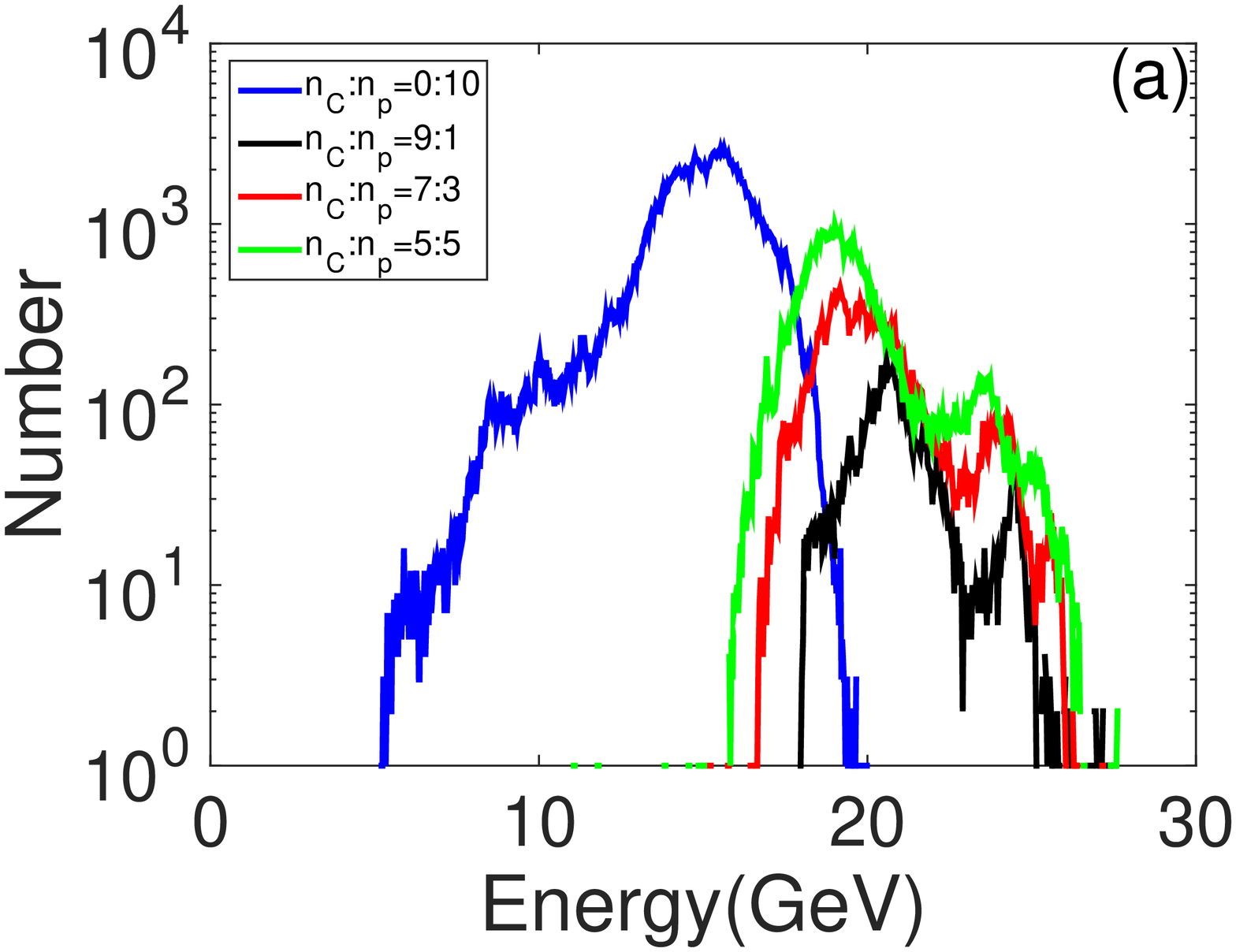}
\includegraphics[width=0.49\textwidth]{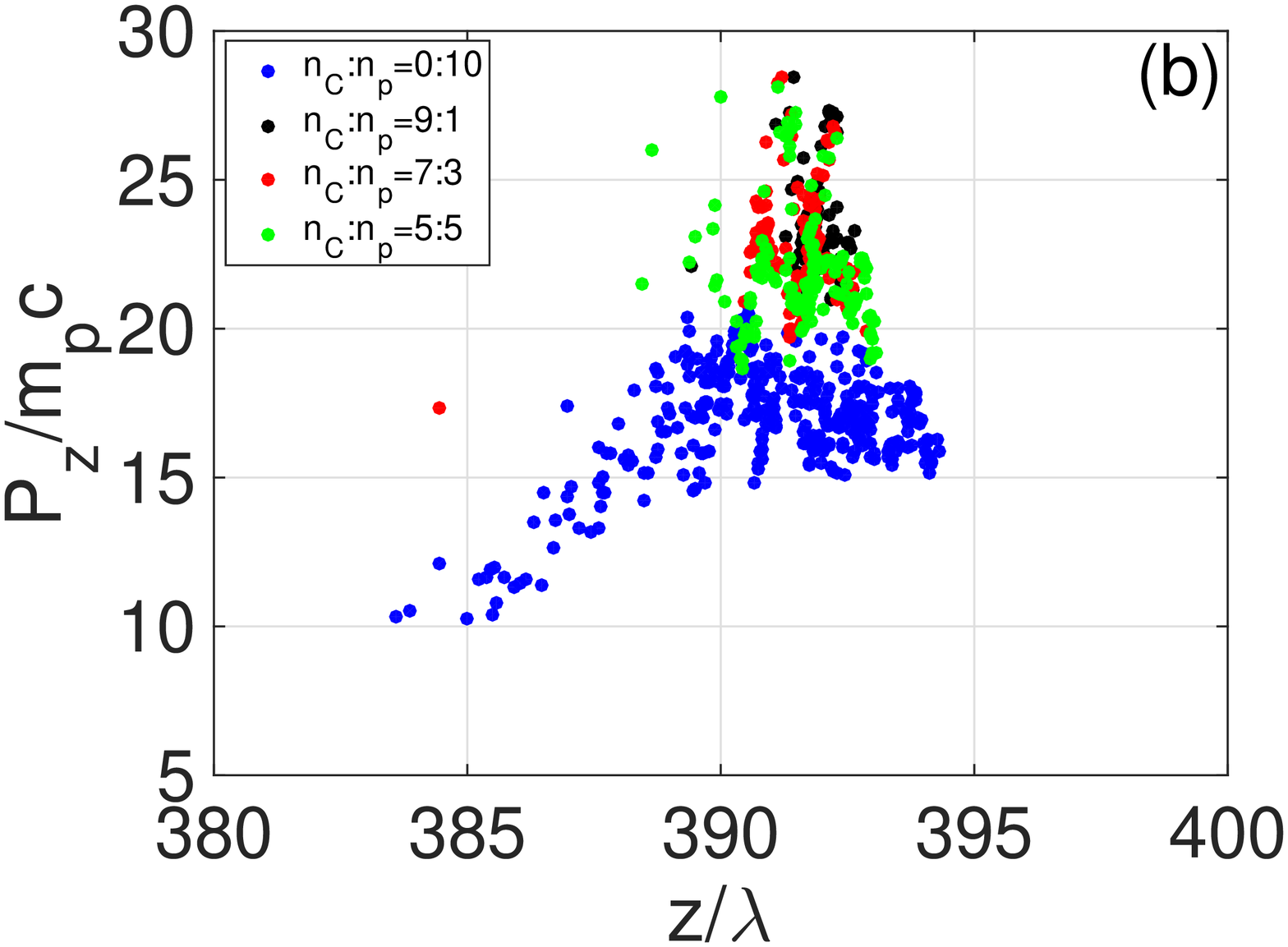}
\caption{(Color online) The energy spectrum of protons is shown in (a) and proton phase space is shown in (b) at $t=400\tau$, when the acceleration gets saturated in all cases. All parameters are the same as for Fig.\ref{fig.2}. }
\label{fig.3}
\end{figure}

Fig.\ref{fig.3} shows the energy spectrum and the phase space of protons at $t=400\tau$, when the LWFA stage ends for all cases. 
In Fig.\ref{fig.3} (a), the energy spectra of different targets clearly show the optimization effects by using CH targets. Not only are the proton maximum energies increased, but also their energy spreads are decreased. The full-width at half-maximum (FWHM) of pure H case is about $18\%$, while that of the CH cases is about $10\%$ for $n_C:n_p=5:5$ case, $7\%$ for $n_C:n_p=7:3$ case and $5\%$ for $n_C:n_p=9:1$ case, respectively. All of above are smaller than that for pure H case.

In Fig.\ref{fig.3} (b), the phase space of the ultra-energetic protons shows the similar results as that in Fig.\ref{fig.3} (a).
Despite a larger number of protons in pure H target, the beam structure is quite spread in space and the maximum energy is only about $20$ GeV. On the contrary, the beam structures from the CH targets are very compact and almost all of the maximum energies are larger than $27$ GeV. Thus, increasing the proton maximum energy by more than $35\%$. 
This result is in accordance with that in Fig.\ref{fig.2}. 
More importantly, it is these compact beam structures of the CH target that lead to the improvement of the monochromaticity. 

\section{Explanations and Discussions}

\begin{figure}[!htb]
\centering
\includegraphics[width=0.49\textwidth]{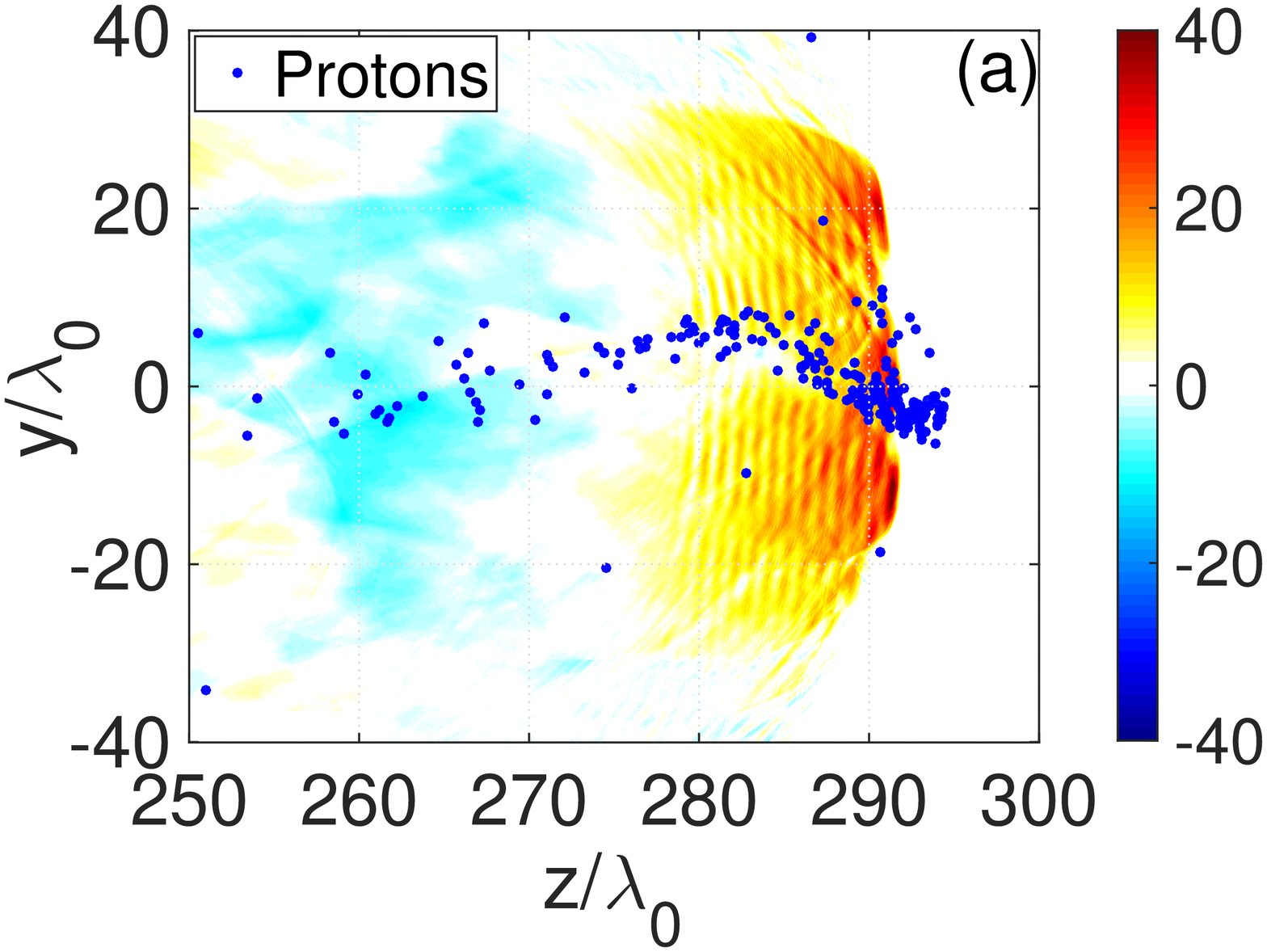}
\includegraphics[width=0.49\textwidth]{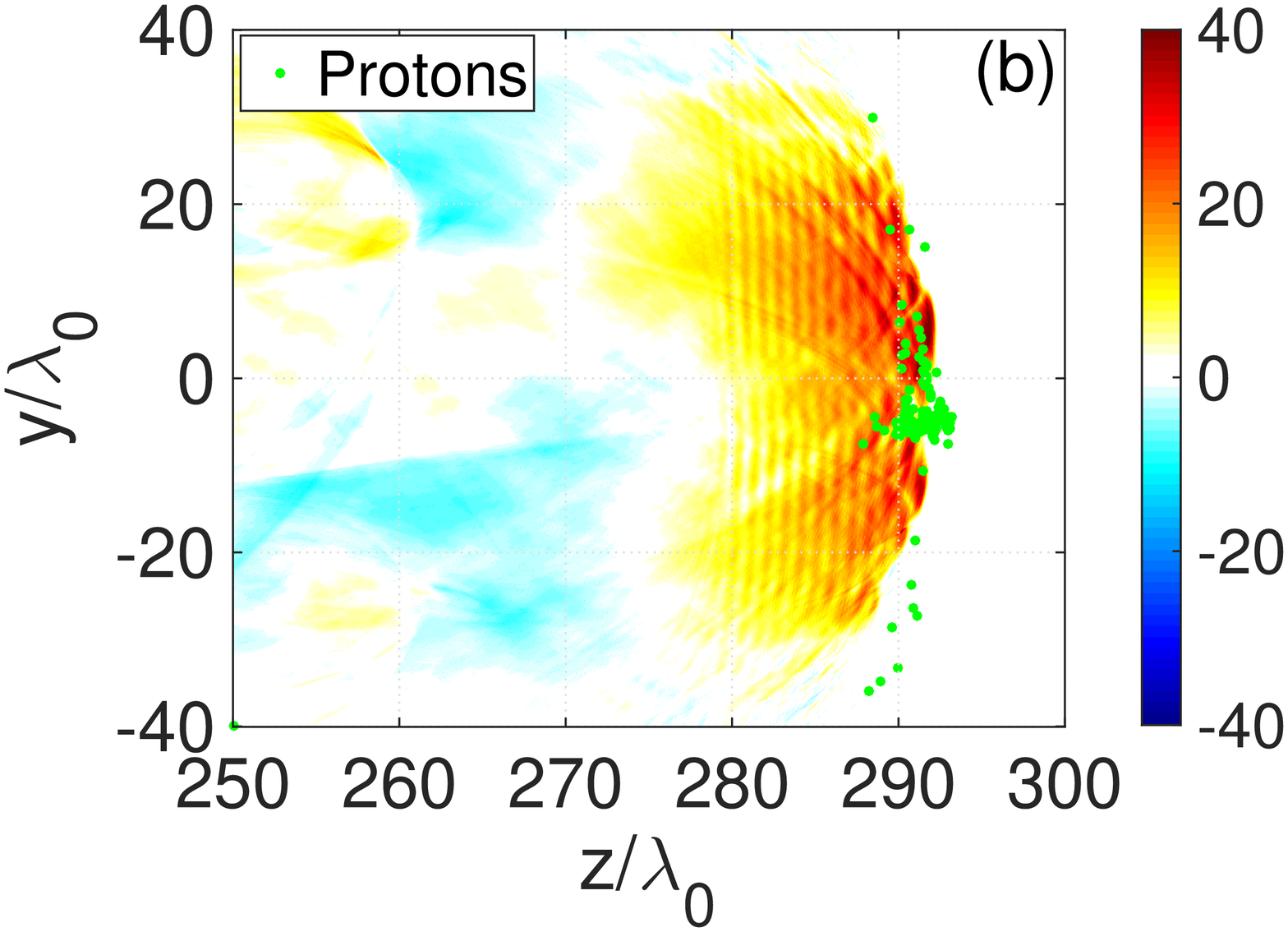}
\caption{(Color online) The normalized longitudinal electrostatic fields $E_z$ for (a) pure H target and (b) $n_C:n_p=5:5$ CH case at $t=300\tau$. Mass-limited target proton positions are dotted in blue for the pure H target and in green for the CH case. All parameters are the same as for Fig.\ref{fig.2}. }
\label{fig.4}
\end{figure}

The normalized longitudinal electrostatic fields shown in Fig. \ref{fig.4} can explain why the use of CH targets can accelerate protons to a higher energy in the LWFA stage. 
As can be seen from Fig. \ref{fig.4} (a), for the pure H case at $t=300\tau$, protons with the highest energy have been accelerated out of the wakefield region. In contrast, for the CH target, most protons are still located at the front of the wakefield, where $E_z$ is the largest, in Fig.\ref{fig.4} (b).
Additionally, it is clearly seen that protons in H target are distributed sparsely in space, resulting in a tail structure with a lower energy behind the most energetic part. While the distribution of the CH protons is very compact with the equal high energy.
When averaging around the region  $-10\lambda < y < 10\lambda$ and $280\lambda < z < 300\lambda$ at $t=300\tau$, there is no doubt that CH protons will get a higher energy, as compared to the pure H target protons.

The longitudinal electrostatic field is also improved in CH target, as compared to the H case. As can be seen from Fig.\ref{fig.4} (a), $E_z$ is smaller in size and weaker in strength, as compared to that in Fig.\ref{fig.4} (b). 
In the pure H target, 
because much of the laser energy is transferred to the relatively low-energy protons, which are distributed in the tail structure of the proton beam, the size and strength of the wakefield are decreased, as compared to those in CH case.
Moreover, when using a CH target, an optimized beam injection effect is realized. The proton beam has a compact structure, without the tail and such relatively low-energy protons in it. 
As a result, the laser energy can be transferred to the trapped protons more effectively.

Similarly, Fig.\ref{fig.4} can also explain why the use of CH target can optimize the proton energy spectrum.
In the positive gradient "debunching" wakefield, protons at the front of the beam will experience a larger longitudinal electrostatic field and will be accelerated to a higher velocity, while those at the back of the beam will experience a smaller longitudinal electrostatic field and a lower speed. Thus, the energy spread of the tail-structured proton beam of H target becomes broader than that of the compact-structured proton beam of CH target.

\begin{figure}[!htb]
\centering
\includegraphics[width=\textwidth]{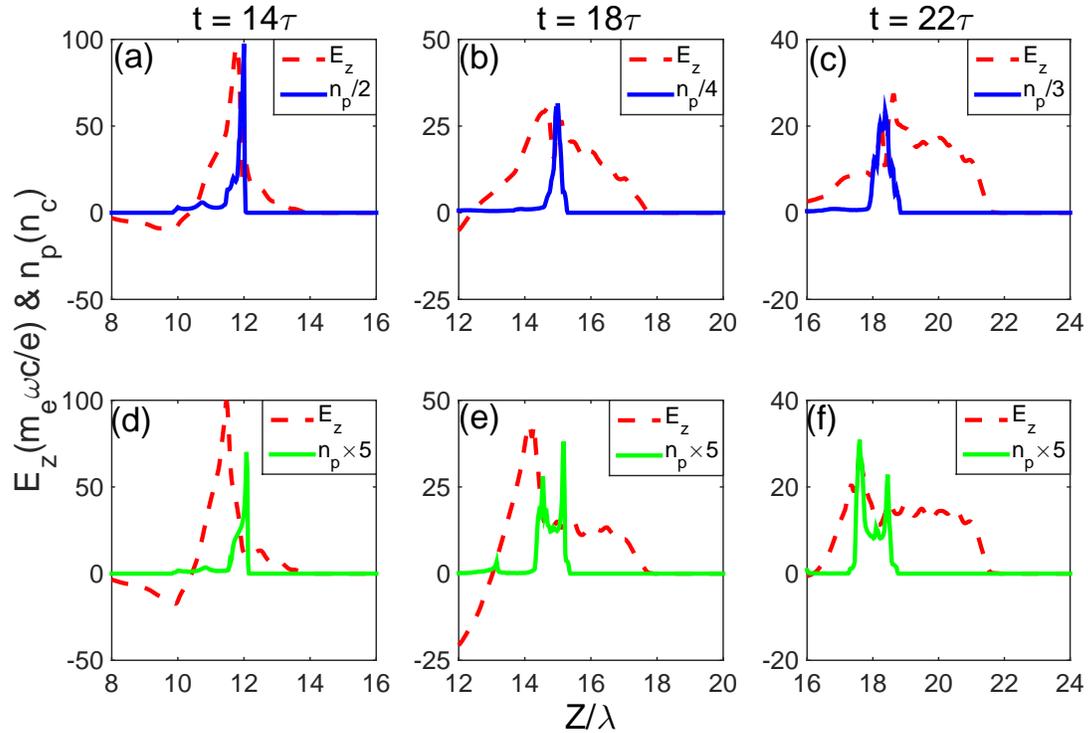}
\caption{(Color online) Snapshots of the normalized longitudinal electrostatic fields $E_z$ (red dashed line) and proton densities for the pure H target (blue solid line) and the $n_C:n_p=5:5$, CH target (green solid line) are presented; (a), (b) and (c) are the pure H case and (d), (e) and (f) are the CH case. Furthermore, (a) and (d) are the result at $t=14\tau$, (b) and (e) are at $t=18\tau$, and (c) and (f) are at $t=22\tau$. All parameters are the same as for Fig.\ref{fig.2}. }
\label{fig.5}
\end{figure}

By comparing the evolution of the electrostatic field in RPDA stage, we can gain further understanding of the physical mechanisms leading to a proton beam quality improvement, when using a CH target instead of a H one.
Fig.\ref{fig.5} shows the snapshots of $E_z$ and $n_p$ at $t=14\tau$, $t=18\tau$ and $t=22\tau$, respectively. Although the value of the $E_z$ for both targets is almost the same at the same time, there is a distinctive difference in the location of the proton beams. 
For CH target, the protons are always located at the negative-gradient part, or the "bunching" part of $E_z$, leading to the compact proton beam structure in Fig.\ref{fig.4} (b), and resulting in a narrower energy spectrum. It is because there are heavy carbon ions behind protons, pushing them with Coulomb repulsion effects. This will also lead to a larger $E_z$, as can be seen in Fig.\ref{fig.5} (e), as compared to that in Fig.\ref{fig.5} (b) at $t=18\tau$.
However, for the pure H target, the protons are located at the "bunching" part of $E_z$ at $t=14\tau$, then at the peak part at $t=18\tau$ and finally at the "debunching" part. 
From then on, the stability of RPDA stage for pure H target becomes lower than that for the CH target. 

In short, our 2D PIC simulation results agree with our former 1D analysis \cite{yao2014} very well, indicating that the target composition scheme has an optimization effect on the combined proton acceleration regime. Moreover, when considering multi-dimensional effects (2D PIC), the optimization scheme performs even better, as compared to 1D scenario. 
Not only are the monochromaticity of the energetic proton beams optimized, but also their maximum energies are improved.    

\section{Parameter Studies}

Aiming at discovering the best set of laser and target parameters, which could improve the maximum energy and monochromaticity of the proton beam, 
parameter studies by varing the CH ratio, the laser intensity and the underdense plasma gas density profile have been done.

\subsection{Effects of Number Density Ratio of CH Target}

\begin{figure}[!htb]
\centering
\includegraphics[width=0.7\textwidth]{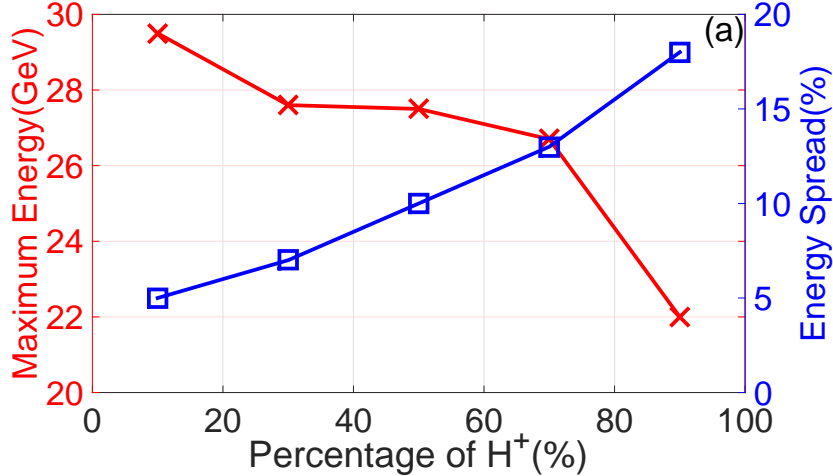}
\caption{(Color online) The scaling of the maximum energy (red line) and energy spread (blue line) of the accelerated proton beams with the different percentage of the hydrogen in the CH target. The data are calculated at $t=400\tau$. All other parameters are the same as for Fig.\ref{fig.2}. }
\label{fig.6}
\end{figure}

Fig.\ref{fig.6} shows the influence of a number density ratio of carbon-to-hydrogen on the quality of the proton beam in the combined proton acceleration regime. By a series of 2D PIC simulations with different density ratio of carbon-to-hydrogen, we found that, as the percentage of number density of $H^{+}$ increases, the maximum energy of the protons is decreasing, while the energy spread of the proton beam is increasing. For $n_C:n_p=1:9$ case, i.e., when the percentage of $H^{+}$ gets $90\%$, the maximum energy is only about $22$ GeV and the energy spread is about $18\%$. However, for $n_C:n_p=9:1$ case, or the percentage of $H^{+}$ is only $10\%$, the maximum energy is almost $30$ GeV and the energy spread is limited to only $5\%$. The result clearly indicates that using a hydrocarbon target in the combined proton acceleration regime can largely improve the acceleration quality. 
   
\subsection{Effects of Laser Amplitude}

\begin{figure}[!htb]
\centering
\includegraphics[width=0.7\textwidth]{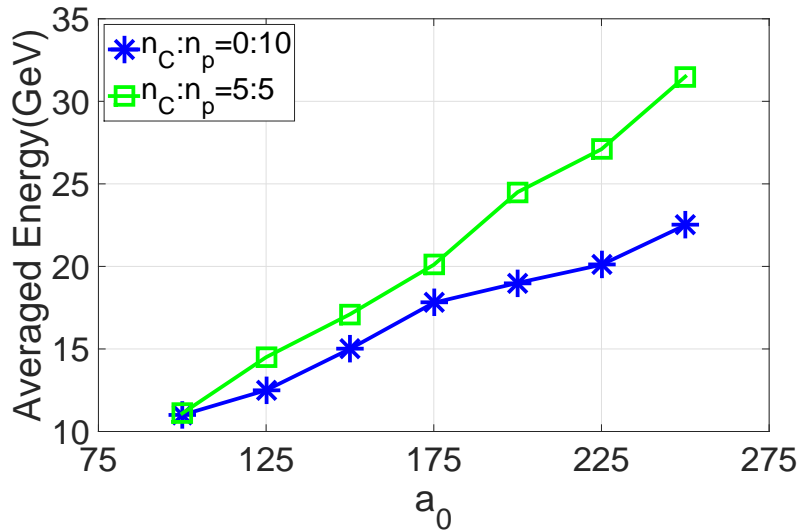}
\caption{(Color online) The scaling of the maximum energy of the accelerated proton beams with the laser intensity. The green line is for the CH target case and the blue line is for the H target case. The data are calculated at $t=400\tau$. All other parameters are the same as for Fig.\ref{fig.2}. }
\label{fig.7}
\end{figure}

In addition to the CH ratio, here we also investigated the influence of laser intensity on the quality of the proton beam in the combined proton acceleration regime. As shown in Fig.\ref{fig.7}, by a series of 2D PIC simulations within a given laser intensity range (from $a_0=100$ to $a_0=250$) , the averaged proton energy is also increased, which is easy to understand. What interests us is that within above laser intensity range, the averaged proton energy of the CH target case is always larger than that of the H target case. 
The above results indicate that, if we want to accelerate the protons to a certain energy, we can use a lower laser intensity by replacing the H target with a CH one. On the other hand, at a fixed laser intensity, by using a CH target instead of a H one, we could get a higher proton energy.

\subsection{Effects of Underdense Plasma Gas Density Profile}
At last, the influence of underdense plasma gas density profile on the proton beam quality is also investigated. 
In the electron LWFA regime, the dephasing length of the trapped electrons can be much longer by employing a positive density gradient\cite{wen2010}. Inspired by this improvement scheme, we can expect to improve the proton energy in the combined proton acceleration regime by using a negative gradient of the underdense background plasma (BP) gas.

\begin{figure}[!htb]
\centering
\includegraphics[width=0.49\textwidth]{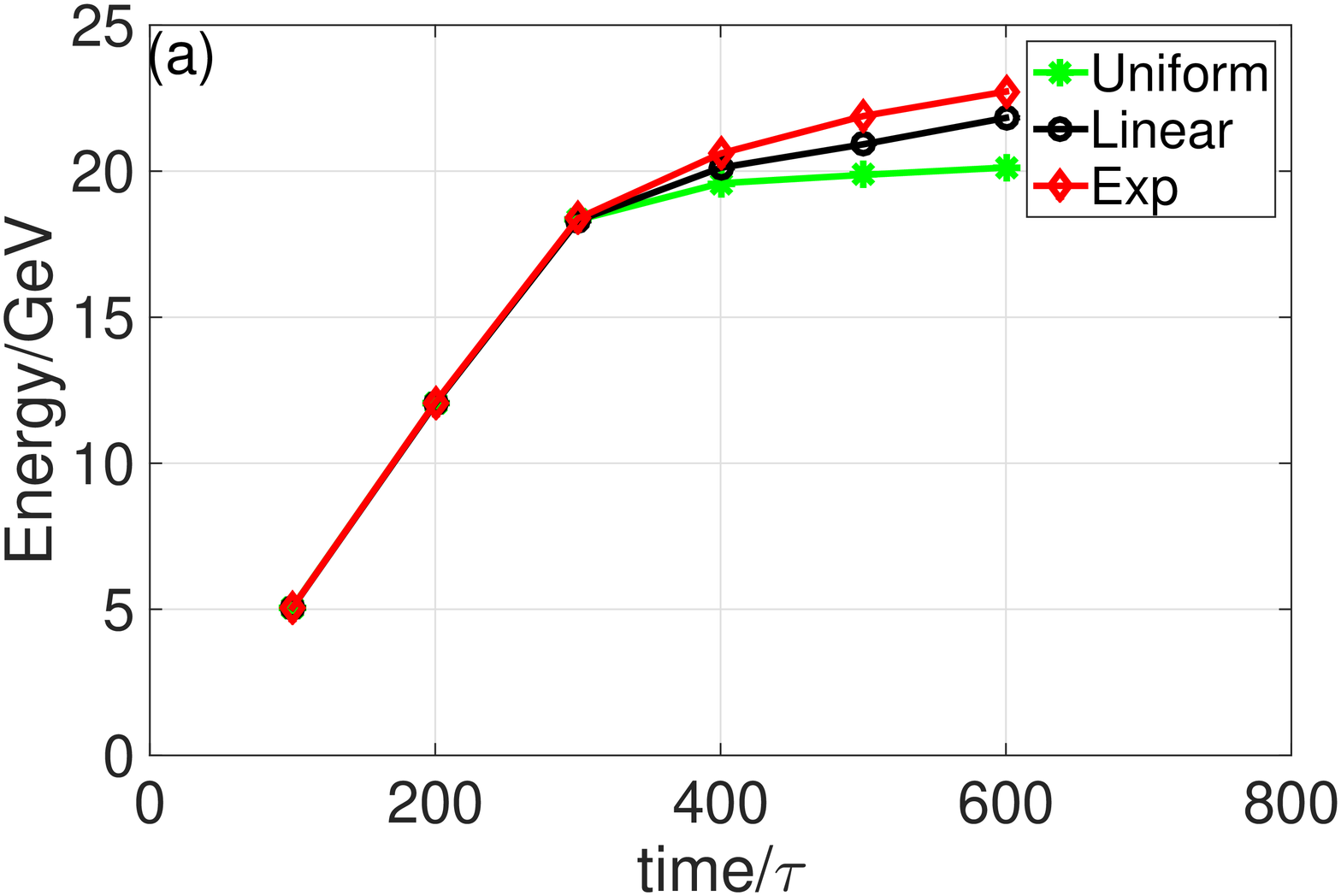}
\includegraphics[width=0.49\textwidth]{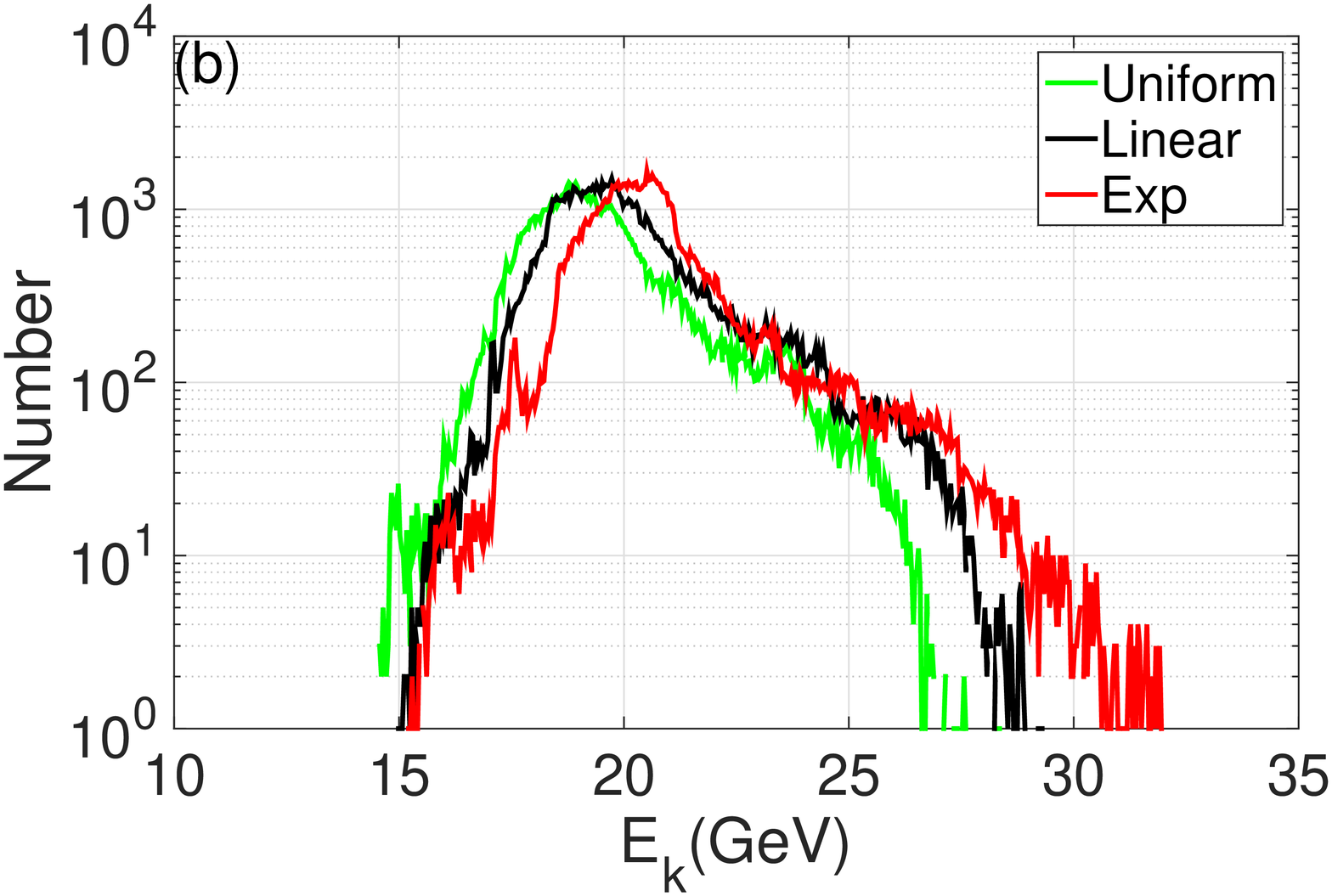}
\caption{(Color online)  (a) The average proton energy with time for CH target for different BP, and (b) the energy spectrum of CH target proton for different BP, the data are calculated at $t=600\tau$. The green solid line is for the uniform BP case, while the black and red solid line are for the linearly and exponentially decreasing gradient BP case, respectively. All other parameters are the same as for Fig.\ref{fig.2}. }
\label{fig.8}
\end{figure}

Fig. \ref{fig.8} (a) shows the average proton beam energy for different underdense background plasma (BP) profiles. A linearly ($n_e/n_c = - 0.0006(z-200)+0.25$) and an exponentially ($n_e/n_c = 0.25\exp{\left[ -(z-200)/125 \right]}$) decreasing background plasma density are used to compare with the uniform case ($n_e/n_c = 0.25$). 
As we expected, using a negative gradient of a background plasma can improve the average proton energy in the combined proton acceleration regime. And larger negative gradient results in a higher proton energy. 
From the energy spectrum of proton beams in Fig. \ref{fig.8} (b), the same result can be seen apparently. 

These results can be explained by analyzing the details of above acceleration carefully.
For the uniform case, the trapped protons will catch up with the accelerating field at the laser front at about $t=400\tau$, and then these protons will stop acceleration and their energy will get saturated. However, for the negative gradient case, the phase velocity of the accelerating field becomes higher than before when it  comes into the low density region. This theory has been well established in a number of former works \cite{zheng2012, bake2012, decker1996} and the velocity can be estimated as, $v_f \approx 1-\frac{3}{2}\frac{n_e}{\gamma_a n_c} $, and $\gamma_a = \sqrt{1+a_0^2}$.
As the accelerating field moves faster, the trapped protons will stay in it for a longer time, and their energy will keep growing without getting saturated. 

Finally, it is worth mentioning that in the electron LWFA regime, as well as, in combined proton acceleration regime, the collimation quality is, as important as, the monochromaticity. However, the long acceleration distance in these regimes make it difficult to keep the energetic protons well collimated. Moreover, laser-plasma instabilities and filamentation will also affect the transport of a laser pulse in the underdense background plasma. To improve the collimation quality, a lot of attempts have been made, including the famous plasma channeling scheme \cite{leemans2006} in the electron LWFA regime, and the transversely modulated background plasma density scheme \cite{zheng2013}, as well as, the LG laser pulse scheme \cite{xmzhang2014} in the combined proton acceleration regime. Efforts of employing a channel-guided scheme in the combined proton acceleration regime have been made by our group and the result will be presented in future publications.   

\section{Summary}

In conclusion, a target composition scheme is proposed, based on 2D PIC simulations, to optimize the maximum energy and energy spectrum of the combined proton acceleration regime (RPDA and LWFA). 
With the use of a CH target, instead of a H one, not only can the proton maximum energy be increased, but also the proton energy spread can be narrowed.  
For example, with the same laser intensity of $1.1\times10^{23}$ $Wcm^{-2}$ and a tritium plasma gas density of $2.8\times10^{20}$ $cm^{-3}$, the use of $n_C:n_e=5:5$ CH target can improve the proton maximum energy by more than $35\%$, from about $20$ GeV to about $27$ GeV. 
Moreover, it can also optimize the energy spread by more than $50\%$ from about $18\%$ down to $7\%$.
Explanations of the acceleration effectiveness and monochromaticity improvement have been made in both the RPDA and LWFA stages. 
From the physical consideration, the carbon ion here plays a key role,
leading to a more stable pre-acceleration of protons and then to a more compact beam structure in the LWFA stage.
Scaling studies of the CH ratio, laser intensity and background plasma density have been investigated in order to further limit the energy spread, to reduce the high requirement for the laser intensity and to increase the proton energy. 

\section{Acknowledgments}
The project was sponsored by the NSFC (Grants No. 11375032, No.11175030)  and the 973 project (Grant No.61319403).
The authors would like to gratefully acknowledge Professor M. M. Skoric, Lihua Cao, Bin Qiao and Xiantu He for their valuable suggestions. And we also thank Fanglan Zheng, Dong Wu, Taiwu Huang and Chengzhuo Xiao for useful discussions.

\bf
\end{document}